\documentclass[prb,aps,twocolumn,floatfix,showpacs,eqsecnum,superscriptaddress]{revtex4}

\usepackage{graphicx,rotating,subfigure,amsmath,amsfonts,amssymb,delarray}
\usepackage{dsfont}
\usepackage[T1]{fontenc}
\usepackage[utf8]{inputenc}
\renewcommand{\vec}[1]{\boldsymbol #1}
\newcommand{\e}{\text{e}}
\newcommand{\im}{\text{i}}
\def\l{\left}
\def\r{\right}
\def\12{\frac{1}{2}}

\newcommand{\be}{\begin{equation}}
\newcommand{\ee}{\end{equation}}
\newcommand{\bea}{\begin{eqnarray}}
\newcommand{\eea}{\end{eqnarray}}

\newcommand{\ud}{\,\mathrm{d}}

\renewcommand\Im{\operatorname{Im}}

\DeclareMathOperator\hside{\theta}

\begin{document}
\bibliographystyle{apsrev} 
\title{Theory of the conductance of interacting quantum wires with good contacts and applications to carbon nanotubes}

\author{Nicholas Sedlmayr}
\affiliation{Department of Physics,
  Technical University Kaiserslautern,
  D-67663 Kaiserslautern, Germany}
\author{Pia Adam}
\affiliation{Department of Physics,
  Technical University Kaiserslautern,
  D-67663 Kaiserslautern, Germany}
\author{Jesko Sirker} 
\affiliation{Department of Physics,
  Technical University Kaiserslautern,
  D-67663 Kaiserslautern, Germany}
\affiliation{Research Center OPTIMAS,
  Technical University Kaiserslautern,
  D-67663 Kaiserslautern, Germany}

\date{\today}

\begin{abstract}
  Using bosonization we derive the dc conductance $G(L,T)$ of an
  interacting quantum wire with good contacts including current
  relaxing backscattering and Umklapp processes. Our result yields the
  dependence of the conductance on length $L$ and temperature $T$ in
  the energy range where the Luttinger model is applicable. For a
  system where only a part of the current is protected by a
  conservation law we surprisingly find an unreduced ideal quantum
  conductance as for a fully ballistic wire. As a second application,
  we calculate the conductance of metallic single-wall carbon
  nanotubes in an energy range where backscattering due to phonons
  dominates. In contrast to previous studies we treat the electrons as
  interacting by using the Luttinger liquid formulation. The obtained
  results for the scaling of the dc conductance with temperature and
  length are compared with experimental data and yield a better
  description than the previously used non-interacting theory.
  Possible reasons for the remaining discrepancies in the temperature
  dependence between theory and experiment are discussed.
\end{abstract}

\pacs{71.10.Pm,73.63.Fg,63.22.Gh}

\maketitle

\section{Introduction}
In a one-dimensional quantum system the quantization of the transverse
momentum means that only a small number of modes are available for
transmission. As a consequence, even a fully ballistic quantum wire
with adiabatic contacts shows a finite {\it quantum
  conductance},\cite{Landauer} $G=2ne^2/h$, where $n$ is the number of
modes and the factor of $2$ arises due to spin degeneracy. If momentum
relaxation by Umklapp or backscattering can be neglected then a wire
with electron-electron interactions can be described as a Luttinger
liquid.  In this case a renormalization of the conductance to
$G=2nKe^2/h$ with a Luttinger parameter $K<1$ for repulsive
interactions might be expected.\cite{KaneFisherPRB,OgataFukuyama}
However, such a renormalization was never observed
experimentally\cite{TaruchaHonda} and does not take into account the
contacts. Assuming that the contacts are adiabatic it has been shown
that the conductance remains unrenormalized by the electron-electron
interactions within the wire.\cite{MaslovStone,SafiSchulz} In any
realistic system the conductance will, however, be affected by
backscattering processes caused, for example, by phonons or impurities
in the bulk of the
wire.\cite{KaneFisherPRB,KaneMeleLee,RistivojevicNattermann} In
addition, backscattering processes at the junctions between the wire
and the leads can also renormalize the
conductance.\cite{FurusakiNagaosa,OgataFukuyama,SedlmayrOhst} Both
contributions can be described at low energies by adding interactions
to the Luttinger liquid Hamiltonian and lead, in general, to a
temperature- and length-dependent conductance.

Experimentally, two types of systems have mainly been used to study
electronic transport in quantum wires. In semiconductor
heterostructures, wide two-dimensional electron gases are formed which
are subsequently laterally confined by applying gate voltages. This
approach makes it possible to obtain very clean wires which show clear
signatures of conductance quantization.\cite{TaruchaHonda} The other
important experimental system are carbon nanotubes
(CNTs).\cite{CharlierBlase} Single-wall carbon nanotubes can either be
semi-conducting or metallic depending on their wrapping vector. The
basic electronic properties of CNTs can be understood by viewing them
as rolled-up graphene sheets.\cite{KaneBalents,EggerGogolin} The
momentum transverse to the tube direction is quantized leading to a
finite number of bands.  For a metallic tube two of these bands cross
at the two Fermi points so that in the low-energy limit we are left
with a system consisting of two left- and two right-moving fermionic
modes. Including spin degeneracy the quantum conductance of a
ballistic CNT would therefore be $G_0=4e^2/h$. In a first important
experiment,\cite{BockrathCobden} ropes of single-wall carbon nanotubes
on Si/SiO$_2$ substrates were studied. In this system charging effects
and resonant tunneling were observed. The conductance was found to be
small and transport dominated by the probability to tunnel an electron
between the contact and the CNT. For a Luttinger liquid with $n=2$
bands such as a CNT, the temperature-dependence of the tunnel
conductance is well-known and scales as $T^\alpha$ with
$\alpha=(K_{c+}+K_{c+}^{-1}-2)/8$ for tunneling into the bulk of the
wire and $\alpha=(K_{c+}^{-1}-1)/4$ for tunneling at an end of the
wire.\cite{EggerGogolin} Experimental
data\cite{BockrathCobden,Bockrath,YaoPostma} for tunneling into the
bulk or an end of a CNT were consistent with these two formulas with a
single Luttinger parameter $K_{c+}\approx 0.2-0.4$ for the total
charge mode as expected by theory.\cite{EggerGogolin,YoshiokaOdintsov}
Spin-lattice relaxation rates $1/T_1$ have also been interpreted in
terms of a Luttinger liquid with $K_{c+}\approx
0.2$.\cite{SingerWzietek,DoraGulacsi}

In recent years, single CNTs have been successfully contacted as well.
In one such experiment,\cite{PurewalHong} almost perfect contacts have
been realized so that the conductance of short wires at low
temperatures was close to the ideal quantum conductance $G_0$.
Contacting the same CNT at various distances Purewal {\it et
  al.}\cite{PurewalHong} were further able to measure the conductance
not only as a function of temperature but also as a function of
length. In contrast to the earlier experiment by Bockrath {\it et
  al.}\cite{BockrathCobden} the good contacts imply that the
conductance is no longer dominated by tunneling processes but rather
by electron backscattering in the bulk of the wire.  Backscattering
can be caused by impurities which are often relevant perturbations and
can completely suppress the conductance below a temperature scale
$T_K$.\cite{KaneFisherPRB} At temperatures $T\gg T_K$, on the other
hand, irrelevant backscattering due to phonons is expected to be the
dominant contribution in clean samples.  Treating the electrons in the
CNT as non-interacting it has been shown that acoustic phonon modes
give rise to a resistivity which increases linearly with
temperature.\cite{KaneMeleLee,ParkRosenblatt,IlaniMcEuen} At
half-filling purely electronic Umklapp scattering also has to be taken
into account and induces a charge gap $\Delta_c$. The conductance for
$T<\Delta_c$ then shows thermally activated behavior.  For $T\gg
\Delta_c$, on the other hand, Umklapp scattering can be treated
perturbatively and leads to a resistivity which---similarly to the
phonon contribution---increases linearly with temperature if the
electrons are assumed to be noninteracting.\cite{BalentsFisher}
However, while the phonon contribution is proportional to $1/R_a$,
where $R_a$ is the tube radius, the Umklapp contribution scales as
$1/R_a^2$ and can therefore be neglected except for very narrow tubes.

In this paper we will consider electronic transport in a quantum wire
described by the generic low-energy Hamiltonian\cite{GiamarchiBook}
\begin{equation}
\label{HamFerm}
H=-\im v_F \sum_{r,\alpha ,\sigma} r\int dx\, \Psi^\dagger_{r\alpha\sigma}(x)\partial_x\Psi_{r\alpha\sigma}(x) \, .
\end{equation}
Here $v_F$ is the Fermi velocity and
$\Psi^{(\dagger)}_{r\alpha\sigma}$ a fermionic annihilation (creation)
operator with $\sigma=\pm$ being the spin, and $\alpha=1,\cdots,n$ a
band index. The fermionic field in the low-energy limit is split into
right movers $\Psi_{+,\alpha\sigma}\equiv\Psi_{R,\alpha\sigma}$ and
left movers $\Psi_{-,\alpha\sigma}\equiv\Psi_{L,\alpha\sigma}$. We
will then use standard Abelian bosonization to express the fermionic
operators in terms of bosonic fields
\begin{equation}
\label{Bos}
\Psi_{r\alpha\sigma}(x)=\frac{\eta_{r\alpha\sigma}}{\sqrt{2\pi \bar a}}\e^{i[k_F(r,\alpha)x+r\sqrt{2\pi}\phi_{r\alpha\sigma}(x)]}.
\end{equation}
Here $\eta_{r\alpha\sigma}$ are Klein factors ensuring the fermionic
commutation rules and $\bar a$ is a cutoff of the order of the lattice
constant $a$. The Fermi momentum $k_F(r,\alpha)$ associated with each
of the modes depends on the band structure of the microscopic model.
We further assume that we can represent the Hamiltonian
(\ref{HamFerm}) using (\ref{Bos}) by
\begin{equation}
\label{HamBos}
H=\frac{1}{2}\sum_{j,\delta}\int dx \left\{v_{j\delta}K_{j\delta}(\partial_x\theta_{j\delta})^2+\frac{v_{j\delta}}{K_{j\delta}}(\partial_x\phi_{j\delta})^2\right\} 
\end{equation}
where the new fields fulfill the commutation relation
$[\theta_{j\delta}(x),\phi_{j'\delta'}(x')]=-\frac{i}{2}
\delta_{jj'}\delta_{\delta\delta'}\mbox{sgn}(x-x')$. The indices $j$,
$\delta$ are now indices of the diagonal modes obtained by combining
the bosonic fields $\phi_{r\alpha\sigma}$ in an appropriate way. The
Hamiltonian (\ref{HamBos}) includes the density-density type
electron-electron interactions which lead to a renormalization of the
velocity $v_F\to v_{j\delta}$ and introduce the Luttinger parameters
$K_{j\delta}$.

The charge density can be expressed as
$\rho=e\sqrt{\frac{2n}{\pi}}\partial_x \phi_{\bar j\bar\delta}$
through one of the bosonic fields which we denote by $\phi_{\bar
  j\bar\delta}$. The current density $j(x)$ is then obtained by the
continuity equation
\begin{equation}
\label{Conteq}
\partial_t\rho(x)=-i[\rho(x),H]=-\partial_x j(x)
\end{equation}
leading to 
\begin{equation}
\label{current}
j=-e\sqrt{\frac{2n}{\pi}}\partial_t\phi_{\bar j\bar\delta} \, .
\end{equation}
In the following we will use the short hand notation $\phi\equiv
\phi_{\bar j\bar\delta}$ for the mode related to the electric charge
and current density.

Our paper is organized as follows: In Sec.~\ref{Sec:Cond} we
generalize the approach by Maslov and Stone\cite{MaslovStone} to
derive a formula for the dc conductance of a quantum wire with good
contacts and some form of damping in the bulk of the wire. In
Sec.~\ref{Sec:Damping} we then use a self-energy approach to consider
damping by Umklapp scattering. In particular, we consider the case of
the integrable $XXZ$ model where only part of the current can decay by
Umklapp scattering while the rest is protected by a conservation law.
Furthermore, we study the general case of damping due to
backscattering assisted by other degrees of freedom such as, for
example, phonons. For CNTs all microscopic parameters relevant for
backscattering by phonons are relatively well-known allowing us to
obtain results for the dc conductance which we compare directly to
experiment in Sec.~\ref{Sec:CNT}. In the conclusions,
Sec.~\ref{Sec:Concl}, we summarize our main results and discuss
possible shortcomings of the obtained formulas for the conductance of
CNTs.

\section{Conductance of a wire with contacts and damping}
\label{Sec:Cond}
\begin{figure}
\includegraphics*[width=1.0\columnwidth]{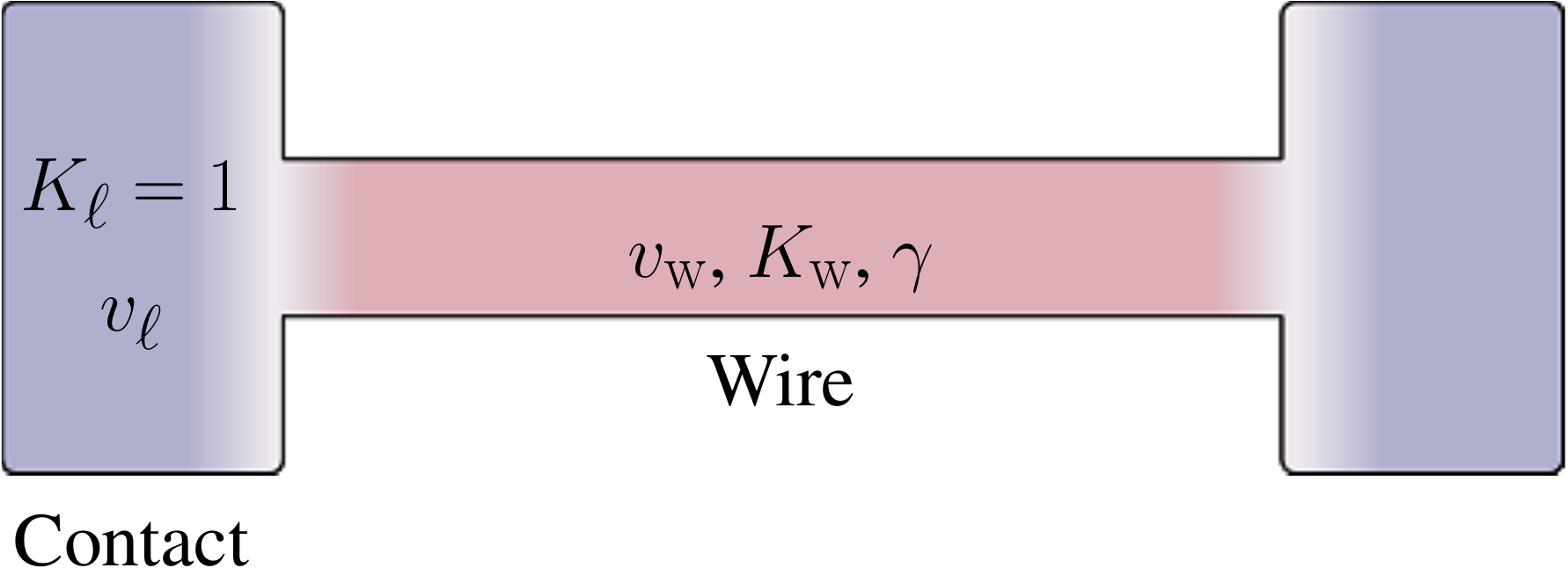}
\caption{(Color online) Schematic view of the end-contacted quantum
  wire. The contacts are assumed to be described by a noninteracting
  channel, $K_\ell=1$, with velocity $v_\ell$. The quantum wire itself
  is a Luttinger liquid with the following parameters for the total
  charge channel: $K_{\mathrm{w}}$ (Luttinger parameter),
  $v_{\mathrm{w}}$ (velocity), and $\gamma$ (current relaxing damping
  rate).}
\label{FigWire}
\end{figure}

We are interested here in the conductance of a finite end-contacted
quantum wire in the linear response regime with some damping in the
bulk of the wire. We assume that we can model the leads as
one-dimensional ballistic channels with Luttinger parameter $K_\ell
=1$. The quantum wire itself, on the other hand, is described by the
Hamiltonian (\ref{HamBos}) with a Luttinger parameter for the total
charge channel $K_{\mathrm{w}}$, a velocity $v_{\mathrm{w}}$, and a damping rate
$\gamma(T)$.  This setup is depicted in Fig.~\ref{FigWire}.
The damping in the wire might stem from electron-electron,
electron-phonon, or electron-impurity interactions. We will consider
the case where the damping is either caused by irrelevant interactions
or cases where the interaction is relevant but we are in a temperature
regime where the renormalized coupling constant for this interaction
is still small so that perturbation theory is applicable. The forward
scattering processes in a one-dimensional conductor are marginal and
take place on a scattering length of the order of the lattice spacing
$a$ leading to new collective excitations described by the Luttinger
liquid Hamiltonian with renormalized values for the velocity $v_{\mathrm{w}}$ and
the Luttinger parameter $K_{\mathrm{w}}$. We will in the following calculate the
conductance in a regime with the backscattering length
$\gamma^{-1}(T)\gg a$ so that we can start from the Luttinger liquid
description of the wire and include the backscattering processes
perturbatively. In addition, we require that $k_B T\ll W_b$ where
$W_b$ is the bandwidth so that a linearization of the dispersion near
the Fermi points is a valid approximation.

There is also one further approximation which we will use in the
following. Strictly speaking one should first calculate the Green's
function for the Luttinger liquid in the contact-wire-contact geometry
following the approach by Maslov and Stone.\cite{MaslovStone} Using
this Green's function one should then perturbatively include the
backscattering to obtain the damping in the system. However, as the
system is no longer homogeneous it seems an impossible task to solve
the resultant Dyson's equation to sum up a series of diagrams.
Instead, we only consider the regime where the coherence length $\xi$
of the considered scattering process is much smaller than the length
$L$ of the wire. These coherence lengths are, for example,
approximately given by $\xi_e\propto\hbar v_F/k_BT$ and
$\xi_{ph}\propto\hbar v_{ph}/k_BT$ for the electron, respectively
phonon, degrees of freedom. Then we can assume that the majority of
the backscattering leading to the damping occurs in the bulk regions
and calculate the relaxation rate $\gamma$ by using the electronic
Green's function for an infinite wire. This damping is then already
included in our approach when calculating the Green's function in the
contact-wire-contact geometry. We will return to this point in Sec.~IV
when comparing with experiments on CNTs. To summarize, we make the
following assumptions: (1) $k_BT\ll W_b$ so that a linearization of
the spectrum is a valid approximation. (2) The backscattering length
$\gamma^{-1}(T)\gg a$ so that we can include forward scattering first
and treat backscattering perturbatively. (3) A coherence length of the
relevant backscattering process $\xi\ll L$ so that most of the
backscattering takes place in the bulk of the wire and can be
calculated by using the bulk electronic Green's function.

Our calculations are based on the Kubo formula which
requires the calculation of the retarded current-current correlation
function. Using the bosonic representation of the current operator,
Eq.~(\ref{current}), the Kubo formula for the conductivity can be
written as\cite{SirkerPereira,SirkerPereira2}
\begin{equation}
\label{Kubo}
\sigma_\omega(x,x')=i\omega\frac{4ne^2}{h}g_\omega(x,x').
\end{equation}
Here the retarded correlation function $g_\omega(x,x')=\langle
\phi(x)\phi(x')\rangle^{\rm ret}_\omega$ can be obtained by an
analytical continuation of the Matsubara function $g_m(x,x')$ at
Matsubara frequencies $\omega_m=2\pi T m$.

Let us first consider the case of an infinitely long wire where
$g$ depends only on $|x-x'|$. In this case the retarded boson
propagator can be expressed in Fourier space as
\begin{equation}
\label{BosonProp}
\langle \phi\phi\rangle_{\rm ret}(q,\omega)=\frac{v_\textrm{w}K_\textrm{w}}{\omega^2-v_\textrm{w}^2q^2-v_\textrm{w}K_\textrm{w}\Sigma_{\rm ret}(q,\omega)} \, .
\end{equation}
The free propagator is obtained by setting the self-energy
$\Sigma_{\rm ret}(q,\omega)=0$. A finite imaginary part of the
self-energy, $\Sigma''_{\rm ret} (q,\omega)\neq 0$, on the other hand,
indicates a finite lifetime of the boson. In writing the boson
propagator in this form we have assumed that a Dyson equation is
valid. In this case the form of the propagator is generic with the
finite lifetime being a consequence of the considered backscattering
process.

According to the assumptions explained above, our starting point for
calculating the conductance of a quantum wire with contacts at both
ends and a damping term in the bulk is the action
\begin{eqnarray}
\label{Action}
S=\frac{1}{2\beta}\sum_{\omega_m}\int dx\,\!\!\!\!\!\!\!&&\bigg\{\frac{\omega_m^2}{v_xK_x}-\frac{\partial}{\partial x}\l(\frac{v_x}{K_x}\partial_x\r) \nonumber \\
&&+2\gamma_x|\omega_m|\bigg\}|\phi(x,\omega_m)|^2
\end{eqnarray}
The position dependent parameters are given by $v_x=v_{\mathrm{w}}$, $K_x=K_{\mathrm{w}}$
and $\gamma_x=\gamma$ in the wire ($0\leq x\leq L$) and by
$v_x=v_\ell$, $K_x=K_\ell$ and $\gamma_x=0$ in the leads ($x<0$ or
$x>L$). $\gamma$ in the wire is calculated using the self energy for a
homogeneous system.

For an infinite quantum wire the action (\ref{Action}) yields after
analytical continuation the boson propagator (\ref{BosonProp}) 
with
\begin{eqnarray}
  \gamma=\frac{1}{2}\Im\frac{\Sigma_{\rm ret}(q=0,\omega)}{\omega}\bigg|_{\omega=0}. 
\end{eqnarray}
Here we have assumed that the self-energy has a regular expansion in
frequency and momentum. Including the real part of the self
energy in lowest order would lead to a renormalization of
$K_{\mathrm{w}}$ and $v_{\mathrm{w}}$ in Eq. (\ref{BosonProp}) but would not
affect the conductance as we will see below.

Let us now come back to the case of a finite quantum wire with
contacts. For the action (\ref{Action}) with abrupt changes in the
parameters the relevant bosonic Green's function
is found by matching across the boundaries between the wire and leads.
We therefore want to solve\cite{SooriSen1,SooriSen2}
\begin{eqnarray}
\label{GF_Eq1}
\left[\frac{\omega_m^2}{ K_x v_x}-\frac{\partial}{\partial x}\left(\frac{v_x}{K_x}\frac{\partial}{\partial x}\right)+2\gamma_x|\omega_m|\right]
g_m(x,x')\qquad
\\ =\delta(x-x').\nonumber
\end{eqnarray}
For convenience we define $\omega_{m\gamma}=\sqrt{\omega^2_m+2\gamma K_{\mathrm{w}}|\omega_m|v_{\mathrm{w}}}$, then we have the free particle equation
\begin{eqnarray}
\left[\frac{\omega_m^2}{ K_\ell v_\ell}-\frac{\partial}{\partial x}\left(\frac{v_\ell}{K_\ell}\frac{\partial}{\partial x}\right)\right]
g_m(x,x')=\delta(x-x')
\end{eqnarray}
for $x$ in the leads and
\begin{eqnarray}
\left[\frac{\omega_{m\gamma}^2}{ K_{\mathrm{w}} v_{\mathrm{w}}}-\frac{\partial}{\partial x}\left(\frac{v_{\mathrm{w}}}{K_{\mathrm{w}}}\frac{\partial}{\partial x}\right)\right]
g_m(x,x') =\delta(x-x')
\end{eqnarray}
for $x$ in the wire. $g_m(x,x')$ is continuous everywhere but there is
a discontinuity in the derivative of $g_m$ at $x=x'$, and
$\lim_{x\to\pm\infty}g_m(x,x')\to0$. The full boundary conditions and
the solution for the Green's function are given in Appendix
\ref{App_GF}.

The current through the device can then be calculated by
\begin{eqnarray}
\label{icurrent}
I=\int_0^L\ud x'\int\frac{\ud\omega}{2\pi}e^{-i\omega t}\sigma_\omega(x,x')E_\omega(x'),
\end{eqnarray}
where $E_\omega(x')$ is the electric field along the wire and the
conductivity $\sigma_\omega(x,x')$ is given by the Kubo formula
(\ref{Kubo}). To determine the current flowing through the device it
is, in general, also necessary to determine how the voltage drops.
For a fully ballistic wire the voltage drop will occur at the two
contacts while the electric field along the wire is zero. For a wire
with damping, however, there will be a voltage drop at the junctions
as well as along the wire. In this case the electric field has to be
determined self-consistently by solving Eq.~(\ref{icurrent}) in
conjunction with the Maxwell equations.  Here we will concentrate on
the dc limit where $\sigma_\omega(x,x')$ becomes position independent
so that the spatial integral in (\ref{icurrent}) can be easily
performed. The current, in this case depends only on the voltage
difference $\Delta V = V_1-V_2$.  The dc conductance then becomes\cite{SooriSen2}
\begin{equation}
\label{dc_cond}
G=\frac{I}{\Delta V} = \frac{2ne^2}{h}\frac{K_\ell}{1+K_{\ell}L\gamma} .
\end{equation}
The resistance consists of two contributions, $R=1/G=R_Q+R_\gamma$.
The quantum resistance $R_Q=\frac{h}{2ne^2}\frac{1}{K_\ell}$ depends
only on the Luttinger parameter $K_\ell$ in the leads whereas the
length and temperature dependent part
$R_\gamma=\frac{h}{2ne^2}L\gamma$ is determined by the damping
$\gamma(K_{\mathrm{w}},T)$ in the bulk of the wire. These two
contributions to the resistance and their general form are expected
and could have been written down without doing any explicit
calculation. What our calculation shows and what is not a priori
clear, however, is that the quantum resistance $R_Q$ is determined
solely by the properties of the leads and is thus not renormalized by
the interactions in the wire while $R_\gamma$ depends only on the
Luttinger parameter $K_{\mathrm{w}}$ and the velocity $v_{\mathrm{w}}$
in the wire. The latter property will break down at temperatures when
the coherence length $\xi$ becomes of the order of the wire length. In
this case the damping $\gamma$ will start to depend on the properties
of both the wire and the contacts.

\section{Damping}
\label{Sec:Damping}
In this section we want to study two examples for damping processes in
the bulk of the wire. First, we consider Umklapp scattering in a
spinless fermion chain at half filling. We explicitly include the
possibility that part of the current is protected by conservation laws
as is the case if the chain is integrable. Second, we consider
backscattering of the electrons due to coupling to other degrees of
freedom. In particular, we discuss backscattering due to interactions
with acoustic or optical phonons.
\subsection{Electron-electron Umklapp scattering}
We want to study this case by considering a concrete microscopic
lattice model given by
\begin{equation}
\label{HXXZ}
H = t \sum_{l=1}^N \big[-\frac{1}{2}\left(c^\dagger_lc^{\phantom\dagger}_{l+1}+ h.c.\right)+\Delta
(n_{l}-\frac{1}{2})(n_{l+1}-{\frac{1}{2}})\big]
\end{equation}
and $n_l=c^\dagger_lc_l$. This so-called $XXZ$ Hamiltonian describes
the hopping of spinless fermions along the chain with hopping
amplitude $t$ and a nearest-neighbor interaction $\Delta$. We want to
concentrate here on the half-filled case $\langle n\rangle =1/2$. In
this case the model is critical for $-1<\Delta\leq 1$ and can be
described at low energies by the Luttinger model, Eq.~(\ref{HamBos}),
with a single mode. The processes which can lead to a relaxation of
the current are due to Umklapp scattering where $2$ electrons are
transferred between the two Fermi points. The leading Umklapp operator
is of the form
$H_U\propto\exp(4ik_Fx)\Psi_R^\dagger(x+a)\Psi^\dagger_R(x)\Psi_L(x+a)\Psi_L(x)+h.c.$
and can be written using the bosonization formula (\ref{Bos}) as
\begin{equation}
\label{Umklapp}
H_U=\lambda\int dx \cos(4\sqrt{\pi}\phi+4k_F x)
\end{equation}
with amplitude $\lambda$, and $k_F=\pi/2$ at half filling. This term
is irrelevant in the regime where the Luttinger model is valid but
becomes marginal for $\Delta=1$ (Luttinger parameter $K=1/2$). We can
therefore calculate the boson propagator (\ref{BosonProp}) in second
order perturbation theory in Umklapp scattering in order to obtain the
damping rate $\gamma$ for this model. Since the $XXZ$ model is
integrable, parameter-free results can be obtained using this approach
and are given in Refs.~\onlinecite{SirkerPereira,SirkerPereira2}. Here
we want to concentrate on one peculiar aspect of the integrability of
the model: there is a quasi-local conservation law\cite{Prosen} which
protects parts of the current from decaying implying an infinite dc
conductivity even at finite temperatures. Within an effective field
theory such a conservation law can be taken into account by using a
memory matrix formalism.\cite{RoschAndrei} The self-energy of the
boson propagator then reads\cite{SirkerPereira2}
\begin{equation}
\label{SelfE_cons}
\Sigma(q=0,\omega)=-\frac{y^{-1}\omega^2}{1+\frac{y^{-1}\omega}{2i\gamma}}\sim\left\{\begin{array}{cc} 
-y^{-1}\omega^2, & y\to\infty \\
-2i\gamma\omega, & y\to 0
\end{array}\right.
\end{equation}
with $y$ describing the overlap between the current operator
$\mathcal{J}=\int dx\, j(x)$ and the conserved quantity $\mathcal{Q}$
\begin{equation}
\label{Overlap}
y=\frac{\langle\mathcal{J}\mathcal{Q}\rangle^2}{\langle\mathcal{J}^2\rangle\langle\mathcal{Q}^2\rangle-\langle\mathcal{J}\mathcal{Q}\rangle^2}.
\end{equation}
The self-energy (\ref{SelfE_cons}) reduces to the standard damping
form if the current is not protected at all ($y\to 0$) whereas the
self-energy vanishes if the current itself is a conserved quantity
($y\to\infty$). For an infinite wire, a partial protection of the
current leads to an infinite dc conductivity in the form of a finite
{\it Drude weight} $D(T)$ appearing in the real part of the
conductivity
\begin{equation}
\label{Drude}
\sigma'(q=0,\omega)=2\pi D(T)\delta(\omega)+\sigma_{\rm reg}(\omega).
\end{equation}

The question we want to address here is what a partial protection of
the current by a conservation law implies for the conductance of a
finite end-contacted wire. In the dc limit the self-energy
(\ref{SelfE_cons}) reduces to $\Sigma(\omega\to 0)\approx
-y^{-1}\omega^2$ if $y\neq 0$. We can now use the same formalism as
before with the only difference that $\omega_{m\gamma}$ is now given
by
\begin{equation}
\label{wmg}
\omega_{m\gamma}=\sqrt{(1+v_{\mathrm{w}}K_{\mathrm{w}}y^{-1})\omega_m^2}.
\end{equation}
From the Green's function in the dc limit given in
Eq.~(\ref{Green_app}) of Appendix \ref{App_GF} we find
\begin{eqnarray}
\label{G_cons}
&&\lim_{\omega\to 0}g(x,x';\omega) = i \frac{K_{\mathrm{w}}}{2\omega\sqrt{1+v_{\mathrm{w}}K_{\mathrm{w}}y^{-1}}}\frac{1+\bar K_{\mathrm{w}}}{1-\bar K_{\mathrm{w}}} \nonumber \\
&=& -i \frac{K_{\mathrm{w}}}{2\omega\sqrt{1+v_{\mathrm{w}}K_{\mathrm{w}}y^{-1}}}\frac{K_\ell}{K_{\mathrm{w}}}\sqrt{1+v_{\mathrm{w}}K_{\mathrm{w}}y^{-1}} \nonumber \\
&=&-i\frac{K_\ell}{2\omega}
\end{eqnarray}
with $\bar K_{\mathrm{w}}$ defined in Eq.~(\ref{barK}). Thus the factor $y$
describing the overlap between the current and the conserved quantity
cancels out as long as $y\neq 0$. Putting (\ref{G_cons}) into the Kubo
formula (\ref{Kubo}) with $2n\equiv 1$ (we have only one mode, $n=1$,
and no spin degeneracy) we find that the dc conductance is
unrenormalized and identical to that of a fully ballistic wire,
$G=\frac{e^2}{h}K_\ell$.

Remarkably, the splitting of the current into a diffusive channel with
a finite relaxation rate $\gamma$ and a ballistic channel, as
described by the self-energy (\ref{SelfE_cons}), does not reduce the
conductance compared to the purely ballistic case. As long as there is
a ballistic channel we find ideal quantum conductance. In some sense
this is analogous to the case of repulsive electron-electron
interactions: the dc conductivity of an infinite wire is reduced by
the Luttinger parameter $K_{\mathrm{w}}$ while the conductance of a finite wire
with Fermi liquid contacts, $K_\ell=1$, remains unchanged. In the
$XXZ$ model considered here the Drude weight is reduced by a factor
$y/(1+y)$ compared to a fully ballistic wire, see
Ref.~\onlinecite{SirkerPereira2}, while we find again that the dc
conductance is not affected. If, on the other hand, $y=0$, i.e. the
transport is purely diffusive and the self-energy is given by
$\Sigma(\omega)=-2i\gamma\omega$, then Umklapp scattering does lead to
a length-dependent conductance as given in Eq.~(\ref{dc_cond}).

Finally, let us remark that we have ignored here any scattering at the
contacts which will, in general, always be relevant and suppress the
conductance at low temperatures. Furthermore, even if the wire is
integrable---something which can only approximately be achieved
experimentally in the sense that integrability breaking terms are
small---integrability will be broken in a setup where the wire is
contacted at its ends. Nevertheless, if we have a wire which is close
to an integrable system then our calculation shows that this can lead
to an anomalously large conductance as long as other scattering, for
example at the contacts, can be ignored.

\subsection{Coupling to other degrees of freedom}
In the previous section we have discussed current relaxation due to
Umklapp scattering caused by electron-electron interactions. In this
case the scattering term (\ref{Umklapp}) involves the transfer of two
electrons from one Fermi point to the other. Backscattering, on the
other hand, where only one electron scatters between the Fermi points,
is kinematically only allowed if the transferred momentum is picked up
by some other degree of freedom of the system. In general, we can
write such an assisted electron backscattering process as
\begin{equation}
\label{coupling}
H_{e-o}=\lambda\int dx\, \hat{O}(x)\sum_{r\alpha\sigma}\Psi^\dagger_{r\alpha\sigma}(x)\Psi_{-r\alpha\sigma}(x).
\end{equation}
where $\lambda$ is the coupling amplitude and $\hat{O}$ the operator
of the other degree of freedom. We will again use the same assumptions
about our system as outlined at the beginning of this section. In
particular, we will calculate the damping rate by using a self-energy
approach with the backscattering electronic Green's function
\begin{eqnarray}
\label{GF_back}
\mathcal G(x,\tau)=-\sum_{r\alpha\sigma}\sum_{r'\alpha'\sigma'}&&\!\!\!\!\!\langle\Psi^\dagger_{r\alpha\sigma}(x,\tau)\Psi_{-r\alpha\sigma}(x,\tau) \\
\!\!\!\!\!\!\!\!\!\!\!\!\!\!&\!\!\!\!\!\!\!\!\!\!\!\!\!\!\times&\!\!\!\!\!\!\!\Psi^\dagger_{r'\alpha'\sigma'}(0,0)\Psi_{-r'\alpha'\sigma'}(0,0)\rangle \nonumber
\end{eqnarray}
calculated for the infinite quantum wire.

We want to concentrate, in particular, on a coupling to the phononic
degrees of freedom. In this case we have $\hat O_q\propto b_q +
b_{-q}^\dagger$ where $b_q$ is the annihilation operator for a phonon
with momentum $q$. In second order perturbation theory in the electron
phonon coupling we then obtain the self-energy
\begin{equation}
  \Sigma(q,\omega_n)\propto\lambda^2\int dx\,d\tau\, D(x,\tau)\mathcal G(x,\tau)[e^{i(qx-\omega_n\tau)}-1] .
\label{Self-E}
\end{equation}
Here $D(x,\tau)$ is a bosonic propagator which in frequency-momentum
space reads
\begin{equation}
\label{phonon_prop}
D(q,\omega_n)=\underbrace{-\frac{2\omega_q}{\omega_n^2+\omega_q^2}}_{\mathcal{D}(q,\omega_n)}d(q)
\end{equation}
where $\mathcal{D}(q,\omega_n)$ is the standard propagator, $\omega_q$
the dispersion relation of the phonon, and $d(q)\sim q$ for an
acoustic phonon while $d(q)\sim\text{const}$ for an optical mode.
Furthermore, $\omega_q=v_{ac}q$ for an acoustic mode where $v_{ac}$ is
the velocity of sound while $\omega_q\approx \mbox{const}$ for an
optical phonon. A Fourier transformation of Eq.~(\ref{phonon_prop})
yields the time-ordered function
\begin{equation}
  D(q,\tau)=-\l[e^{-\omega_q|\tau|}+2n_B(\omega_q)\cosh\omega_q\tau\r]d(q) \, ,
\label{phonon_prop2}
\end{equation}
with $n_B$ being the Bose function, which will be very useful in the
following.

\section{Conductance of single-wall carbon nanotubes}
\label{Sec:CNT}
The conductance of single-wall carbon nanotubes has been measured in
Ref.~\onlinecite{PurewalHong} for a wide range of tube lengths and
temperatures. The increase of the resistance at intermediate
temperatures has been attributed to electron-phonon scattering while
at low temperatures also impurities and localization effects are
expected to play a role.\cite{SundqvistGarcia-Vidal} So far the
resistance of carbon nanotubes due to electron-phonon scattering has
been calculated by taking only one of the acoustic modes into account
and by assuming that the electrons are non-interacting. These
assumptions lead to a resistance which increases linearly with
temperature.\cite{KaneMeleLee} Here we will calculate the resistance
by taking first the electron-electron forward scattering into account
using the Luttinger liquid formalism and then treating the
backscattering due to electron-phonon coupling perturbatively. Note
that the CNTs in experiment\cite{PurewalHong} have lengths $L\sim
10^{-5}-10^{-6}$ m while the coherence length $\xi_e=\hbar
v_{c+}/k_BT\ll L$ for $T> T_c\approx 30$ K so that our approach
outlined in Sec.~\ref{Sec:Cond} is applicable in this temperature
regime.

\subsection{Theoretical results}
At low energies the bosonized Hamiltonian of a single-wall carbon
nanotube including the density-density type interactions is given by
Eq.~(\ref{HamBos}) where $j\delta =c+,c-,s+,s-$ describe the total and
relative parts ($\delta=+,-$) of charge ($j=c$) and spin
($j=s$).\cite{EggerGogolin} The Luttinger parameter of the total
charge mode is given by $K_{c+}\approx 0.2-0.4$ while the other
Luttinger parameters are hardly renormalized, $K_{j\delta\neq
  c+}\approx 1$. Accordingly, the velocity $v_{c+}\approx v_F/K_{c+}$
is larger than the velocities of the other modes, $v_{j\delta\neq
  c+}\approx v_F$. Away from special commensurate fillings---which is
the experimentally relevant case we will concentrate on---the relevant
interactions which are not of density-density type leave the $c+$-mode
unaffected. The other modes become in principle gapped at very low
energies.\cite{EggerGogolin} However, the energy scale where this
happens is of the order of a few Millikelvin.\cite{CNT_LectureNotes}
For the temperature range we are interested in these small gaps do not
play any role and we neglect the interactions responsible for these
gaps in the following.

Carbon nanotubes can be characterized by their wrapping vector
$\vec{C}=n\vec{a}_1+m\vec{a}_2$ where $\vec{a}_{1,2}$ are the basis
vectors of the hexagonal lattice and $n,m$ integer numbers. CNTs with
$(n,m)=(n,0)$ are called zigzag tubes while tubes with $(n,n)$ are
called armchair tubes. The structure of the tube has important
consequences for the electronic structure as well as for the phonon
modes relevant for electron backscattering. Tubes where $n-m$ is zero
or a multiple of three are metallic while other tubes are
semiconducting.  This means, in particular, that all armchair tubes
are metallic. In an armchair tube, only an acoustic mode causing a
twisting of the tube leads to a relaxation of an electric current
while in a zigzag tube an acoustic stretching mode and an optical
breathing mode contribute.\cite{DresselhausEklund} For a generic
$(n,m)$-tube the resistivity due to electron-phonon scattering is thus
given by\cite{SuzuuraAndo}
\begin{equation}
\rho_{e-ph}(T)=\sin^2(3\eta)\rho_t(T) +\cos^2(3\eta)[\rho_s(T)+\rho_B(T)].
\label{Res_CNT}
\end{equation}
Here $\eta$ is the chiral angle with $\eta=0$ for zigzag tubes and
$\eta=\pi/6$ for armchair tubes. $\rho_{t,s,B}(T)$ are the
contributions due to the twiston, stretching, and breathing mode,
respectively. 

We will first focus on the twiston mode. The contribution of this mode
to the resistivity has already been studied by Kane {\it et al.},
Ref.~\onlinecite{KaneMeleLee}, however, in this paper it has been
assumed that the electrons are non-interacting. Here we want to
generalize this calculation by using the bosonized Hamiltonian
(\ref{HamBos}), thus taking the electron-electron interactions into
account as well. The long wavelength twistons can be described by a
continuum theory\cite{KaneMeleLee}
\begin{equation}
H_t=\frac{1}{2}\int dx \l\{M_t(\partial_t\Phi)^2 +C_t(\partial_x\Phi)^2\r\}
\label{Twiston}
\end{equation}
where $M_t$ is the moment of inertia per unit length and $C_t$ the
twist modulus. These parameters are relatively well-known for CNTs and
are given in Appendix \ref{App}, table \ref{CNT_parameters}. The
twiston has dispersion $\omega_q=v_tq$ with velocity
$v_t=\sqrt{C_t/M_t}$. For the specific case of the twiston we might
then write the generic backscattering term (\ref{coupling}) as
\begin{equation}
\label{twist-coup}
H_{e-t}=\lambda_t\int dx\, \partial_x\Phi \sum_{r\alpha\sigma}\Psi^\dagger_{r\alpha\sigma}\Psi_{-r\alpha\sigma}
\end{equation}
where $\lambda_t^2=M_tg_2^2/(\pi R_aM)$ is the electron-twiston
coupling constant expressed in terms of the electron-phonon coupling
constant $g_2$, the radius of the tube $R_a$, and the carbon mass per
unit area $M$.\cite{MartinoEgger} Approximate values for these
parameters are listed in Appendix \ref{App}, table
\ref{CNT_parameters}. Bosonizing the backscattering term leads
to\cite{MartinoEgger,ChenAndreev}
\begin{eqnarray}
\label{back_bos}
 \sum_{r\alpha\sigma}\Psi^\dagger_{r\alpha\sigma}\Psi_{-r\alpha\sigma}\hspace{5.5cm}\\\nonumber
=\frac{4}{\pi\bar{a}}\cos(\sqrt{\pi}\phi_{c+}-2q_Fx)\cos(\sqrt{\pi}\phi_{c-})\prod_{\delta=\pm}\sin(\sqrt{\pi}\phi_{s\delta})\\\nonumber
+\frac{4}{\pi\bar{a}}\sin(\sqrt{\pi}\phi_{c+}-2q_Fx)\sin(\sqrt{\pi}\phi_{c-})\prod_{\delta=\pm}\cos(\sqrt{\pi}\phi_{s\delta})
\end{eqnarray}
where $\bar{a}\approx a$ is a short distance cutoff of the order of
the lattice constant $a$ and $2q_F$ the transferred momentum. In the
following we will set $\bar{a}\equiv a=1$.

We now want to calculate the self-energy, Eq.~(\ref{Self-E}). The
twiston propagator in Matsubara space is given by 
\begin{equation}
  D_t(q,\omega_n)=\underbrace{\frac{q}{2M_tv_t}}_{d(q)}\underbrace{\frac{-2\omega_q}{\omega_n^2+\omega_q^2}}_{\mathcal{D}(q,\omega_n)}.
\label{twiston_prop}
\end{equation}
with the time-ordered function given by Eq.~(\ref{phonon_prop2}). 
Since $v_t\ll v_{c+}$ we can neglect the momentum transferred onto the
phonon. The typical momentum of an electron is $q\sim T/v_{c+}$ so
that $v_tq/T \sim v_t/v_{c+} \ll 1$ which means that the twiston mode
is always heavily populated.\cite{KaneMeleLee} Using these
approximations for Eq.~(\ref{phonon_prop2}) we find
\begin{equation}
  D_t(q,\tau)\approx-\frac{q}{2M_tv_t}\coth\l(\frac{v_tq}{2T}\r)\approx -\frac{T}{C_t}.
\label{twiston_prop3}
\end{equation}
Using the retarded functions the self-energy (\ref{Self-E}) for the
electron-twiston coupling reads
\begin{equation}
  \Sigma^{\rm ret}_t(q,\omega)=2\pi\lambda_t^2\int dt\, [D_t(x=0,t)\mathcal G(x=0,t)]_{\rm ret}(e^{i\omega t}-1)
\label{Self-E2}
\end{equation}
with a retarded Green's function $\mathcal G$ which can be calculated
using the bosonization result (\ref{back_bos}),\cite{Schulz86}
\begin{eqnarray}
  \mathcal G_{\rm ret}(0,t)&=&\hside(t)\frac{4}{\pi^2}\l[\prod_{j,\delta}\l(\frac{\pi T}{v_{j\delta}}\r)^{K_{j\delta}/2}\r]\\
  &&\qquad\times \sin\l(\frac{\pi K}{4}\r) |\sinh(\pi T t)|^{-K/2}\nonumber
\label{Gret}
\end{eqnarray}
with $K=\sum_{j\delta} K_{j\delta}$. Using the formula
\begin{equation}
\int_0^\infty dt\frac{\e^{i\omega t}}{\sinh^{K/2}(\pi T t)} =\frac{2^{K/2-1}}{\pi T} B\l(\frac{K}{4}+\frac{i\omega}{2\pi T},1-\frac{K}{2}\r)
\label{Beta_int}
\end{equation}
where $B(x,y)=\Gamma(x)\Gamma(y)/\Gamma(x+y)$ is the Beta function we
can evaluate the self-energy (\ref{Self-E2}). The integral in
(\ref{Beta_int}) is only convergent if $0<K<2$.  Here we are, however,
only interested in the imaginary part, which is responsible for the
damping, in the limit $\omega/T\to 0$.  This part turns out to be
universal even for $K>2$ while the real part will depend on the cutoff
needed to regularize the integral in this
case.\cite{Schulz86,SirkerPereira2} We can thus expand the Beta
function and find, after reinserting factors of $\hbar$ and lattice
constant $a$ and using $K_{j\delta}=1$ for $j\delta\neq c+$,
\begin{eqnarray}
\label{twiston_relaxation}
  \rho_t(T)&=&R_Q\frac{\Sigma_{\rm ret}''}{2\omega}\bigg|_{\omega=0}\\
  &=&R_Q\lambda_t^2\frac{2^{(3+K_{c+})/2}K_{c+}^{K_{c+}/2}}{\pi a\hbar v_F C_t}\cos\l(\frac{\pi}{4}(3+K_{c+})\r) \nonumber \\
  &\times& B\l(\frac{3+K_{c+}}{4},-\frac{1+K_{c+}}{2}\r)\l(\frac{\pi a k_B T}{\hbar v_F}\r)^{(1+K_{c+})/2} \nonumber
\end{eqnarray}
where $R_Q=\frac{h}{4e^2}$ is the quantum resistance. In the
non-interacting limit this reduces to the known
result\cite{KaneMeleLee} 
\begin{equation}
\label{twiston_relax_nonint}
\rho_t(T)=R_Q\frac{2\lambda_t^2}{C_t v_F^2\hbar^2}k_BT.
\end{equation}
Note that all the parameters in Eq.~(\ref{twiston_relaxation}) are
approximately known, see table \ref{CNT_parameters}.

The stretching mode is also an acoustic mode and the calculation is
analogous to the one for the twist mode. Following
Refs.~\onlinecite{SuzuuraAndo,MartinoEgger} we assume a single energy
scale for the electron-phonon coupling. This allows us to directly
relate the phonon propagators of the two modes
$D_s(q,i\omega_n)=\mathcal{A}\, D_t(q,i\omega_n)$ where
$\mathcal{A}\approx 0.66$ can be expressed in terms of the bulk and
shear modulus of the tube, see table \ref{CNT_parameters}. As a
consequence we have $\rho_s(T)=\mathcal{A}\,\rho_t(T)$.

Finally, we have to consider the optical breathing mode. Its energy
$\hbar \omega_B$ is inversely proportional to the tube radius (see
table \ref{CNT_parameters}) and varies for the systems we are
interested in between $420$ K for a $(10,0)$ tube and $210$ K for a
$(20,0)$ tube. Because these energies are still much smaller than the
bandwidth of a metallic carbon nanotube which is of the order of a few
electron volts we can continue to use the bosonized Hamiltonian. The
self-energy is again calculated starting from Eq.~(\ref{Self-E}) with
the boson propagator defined by
\begin{equation}
D_B(q,\omega_n)=\underbrace{\frac{1}{2\omega_B MR_a^2}}_{d(q)}\underbrace{\frac{-2\omega_B}{\omega_n^2+\omega_B^2}}_{\mathcal{D}_B(q,\omega_n)}
\label{breathing_prop}
\end{equation}
The phonon propagator is momentum independent and the self-energy
reduces, after analytical continuation, to (\ref{Self-E2}) with
$D_t\to D_B$ and $\lambda_t^2\to \lambda_B^2=g_2^2/\pi
R_a$.\cite{MartinoEgger} However, now the time dependence has to be
kept explicitly leading---after reinserting factors of $\hbar,\, k_B,\,
a$---to
\begin{widetext}
\begin{eqnarray}
  \rho_B(T)&=&R_Q\frac{g_2^2}{\pi R_a}\frac{\hbar\omega_B}{2(B+\mu)}\frac{2^{(1+K_{c+})/2}K_{c+}^{K_{c+}/2}}{\pi(\hbar v_F)^2} 
\l(\frac{\pi a k_B T}{\hbar v_F}\r)^{(K_{c+}-1)/2} \\
  &\times & \l[\cos\l(\frac{\pi(3+K_{c+})}{4}\r)\mbox{Im}F\l(\frac{\hbar\omega_B}{2\pi k_B T}\r)+(1+2n_B(\omega_B))\sin\l(\frac{\pi(3+K_{c+})}{4}\r)\mbox{Re}F\l(\frac{\hbar\omega_B}{2\pi k_B T}\r)\r] \nonumber
\label{breath_relaxation}
\end{eqnarray}
with 
\begin{equation}
F(x)=\frac{\Gamma\l(1-\frac{3+K_{c+}}{2}\r)\Gamma\l(\frac{3+K_{c+}}{4}-ix\r)}{\Gamma\l(1-\frac{3+K_{c+}}{4}-ix\r)}\l[\Psi_0\l(1-\frac{3+K_{c+}}{4}-ix\r)-\Psi_0\l(\frac{3+K_{c+}}{4}-ix\r)\r]
\label{F_eqn}
\end{equation}
\end{widetext}
where $n_B(\omega)$ is the Bose distribution function, $\Psi_0(x)$ the
Digamma function, and $B$ and $\mu$ the bulk and shear modulus
respectively. We have also made use of the relation
$\omega_B^2=(B+\mu)/MR_a^2$.\cite{SuzuuraAndo}


In total, we have obtained a result for the resistivity
(\ref{Res_CNT}) of a carbon nanotube due to phonon assisted
backscattering which includes the electron-electron interactions of
density-density type and depends only on microscopic parameters which
can be theoretically estimated or measured experimentally. It is
important to note that the prefactor of the two acoustic modes is
inversely proportional to the radius $R_a$ of the tube so that this
scattering process becomes less important the wider the tube is. The
prefactor of the breathing mode, on the other hand, scales as
$1/R_a^2$. At the same time, however, the energy of the breathing mode
decreases as $\hbar \omega_B\sim 1/R_a$. The breathing mode
contribution (\ref{breath_relaxation}) consists of two parts: A part
describing the absorption of thermally excited phonons, and a part
describing spontaneous emission of a phonon. Both parts are
exponentially suppressed at low temperatures.

\subsection{Numerical evaluation and comparison with experiment}
\begin{figure}
\includegraphics*[width=1.0\columnwidth]{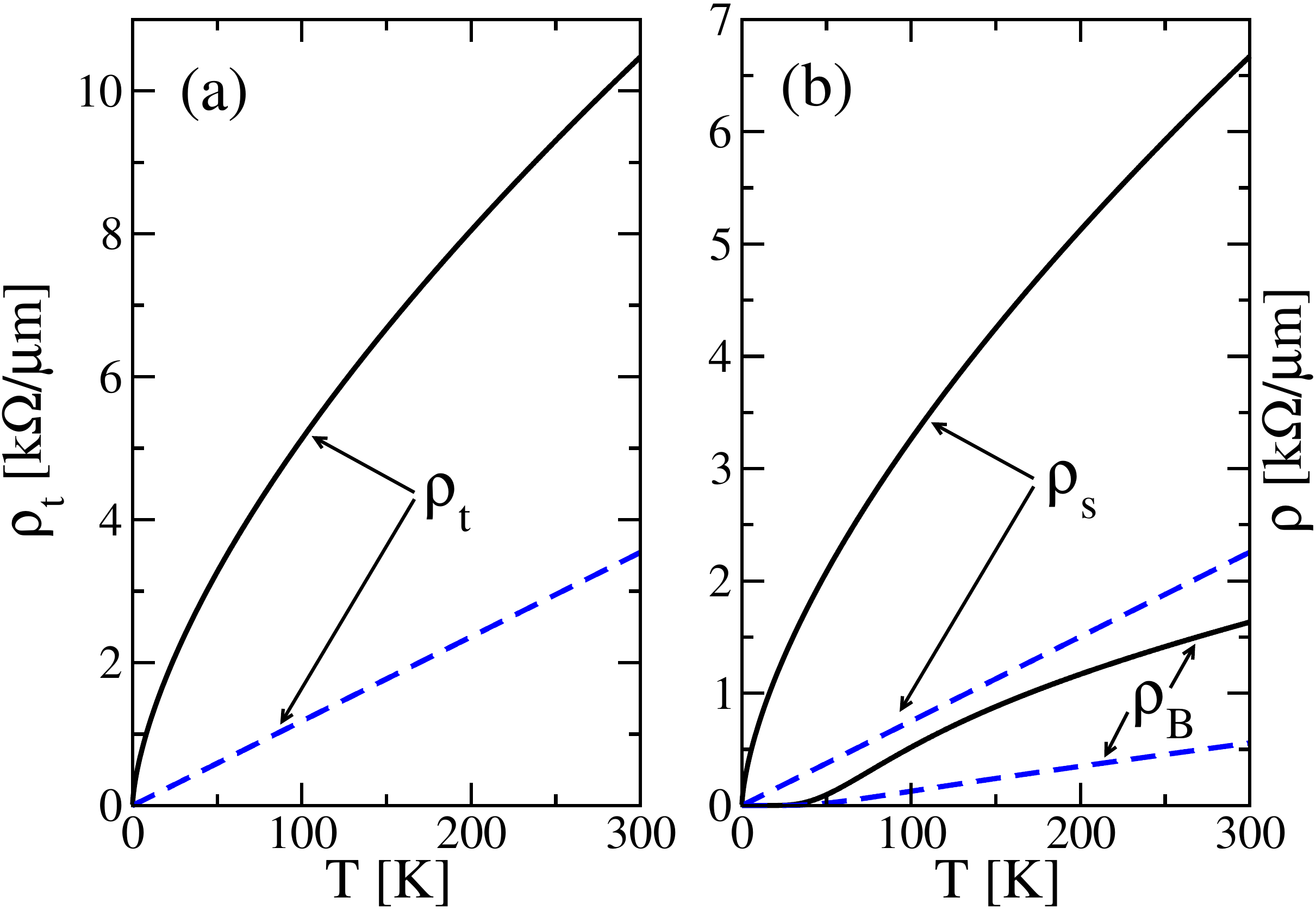}
\caption{(Color online) (a) Resistivity $\rho_t$ of a $(10,10)$
  armchair tube caused by coupling to the twiston mode in the
  noninteracting case ($K_{c+}=1$, blue dashed line) and in the
  interacting case ($K_{c+}=0.3$, black solid line) with $g_2=1.5$ eV.
  (b) $\rho_s$ and $\rho_B$ for a $(18,0)$ zigzag tube in the
  noninteracting case ($K_{c+}=1$, blue dashed line) and in the
  interacting case ($K_{c+}=0.3$, black solid line) with $g_2=1.5$ eV
  and all other parameters as given in the Appendix \ref{App}, table
  \ref{CNT_parameters}.}
\label{Res_examples}
\end{figure}

We can now evaluate (\ref{Res_CNT}) for different tubes. As examples,
we show in Fig.~\ref{Res_examples}(a) the result for a $(10,10)$
armchair tube. In this case the chiral angle is $\eta=\pi/6$ and only
the twiston mode (\ref{twiston_relaxation}) contributes. In
Fig.~\ref{Res_examples}(b) the resistivity of a $(18,0)$ zigzag tube
is shown where $\eta=0$. The stretching mode contribution,
$\rho_s(T)=\mathcal A\rho_t(T)$, and the breathing mode contribution
(\ref{breath_relaxation}) to the resistivity are shown separately.
Both tubes have a comparable radius of $R_a\approx
7\,\mathring{\mathrm{A}}$ and all other parameters are as given in
table \ref{CNT_parameters} of Appendix \ref{App}. In both cases we
have set $K_{c+}=0.3$.
The linear resistivity of the acoustic modes changes into $\rho\sim
T^{(1+K_{c+})/2}$ which for $K_{c+}<1$ leads to a negative curvature
as a function of temperature. Furthermore, the interactions
substantially increase the resistivity by about a factor of $3$ at
room temperature in the examples shown in Fig.~\ref{Res_examples}. The
contribution of the breathing mode for the zigzag tube is
significantly increased by electron-electron interactions as well.

\begin{figure}
\includegraphics*[width=1.0\columnwidth]{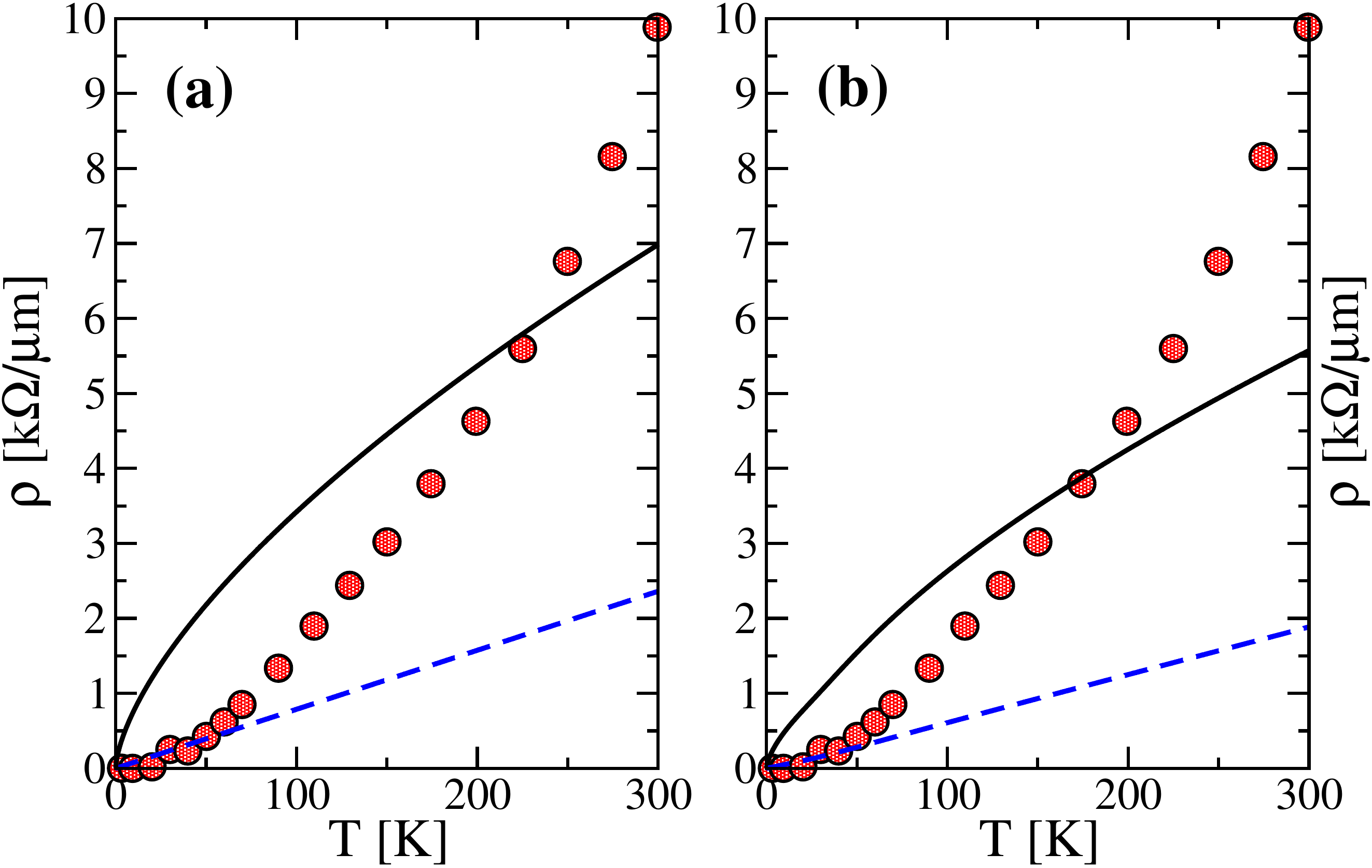}
\caption{(Color online) Experimental data taken from
  Ref.~\onlinecite{PurewalHong} (dots) compared to theoretical results
  for (a) a $(15,15)$ armchair and (b) a $(27,0)$ zigzag tube.
  Results with $g_2=1.5$ eV in the non-interacting case $K_{c+}=1$
  (blue dashed lines) and in the interacting case with $K_{c+}=0.3$
  (black solid lines) are shown.}
\label{Exp_data}
\end{figure}

For a comparison with experiment we will concentrate here on the
results obtained in Ref.~\onlinecite{PurewalHong}. In this work
single-wall carbon nanotubes with lengths up to $1$ mm were deposited
on a Si/SiO$_2$ substrate and the resistance of a {\it single}
tube as a function of temperature and length was studied.  Except for
low temperatures where impurities and possibly localization effects
play an important role, the resistance was found to scale linearly
with length. We will focus here on device `M1' from
Ref.~\onlinecite{PurewalHong} which seems to have the cleanest
contacts and is metallic down to low temperatures. The resistivity of
this device as a function of temperature is shown in
Fig.~\ref{Exp_data} where we have subtracted a small constant
contribution due to non-ideal contacts and impurities.
While scattering at impurities and at the contacts is expected to be
relevant and therefore lead to an increasing resistivity for $T\to 0$
at temperatures $T\ll T_K$ this temperature regime has not been
reached in experiment and taking this contribution as constant seems
to be a reasonable approximation. The diameter of the tube in device
`M1' has been determined to be $d\approx 2$ nm. The chiral angle,
however, is not known. If the tube would be an armchair tube, then
this diameter is consistent with a $(15,15)$ tube while a zigzag tube
of this diameter would be close to a $(27,0)$ tube. We thus plot in
Fig.~\ref{Exp_data} the theoretical results for both kinds of tubes in
comparison to the experimental data. 

Including the electron-electron interactions substantially increases
the resistivity and leads to a much better overall quantitative
agreement with experiment. This is, in particular, true if we assume
an armchair tube, Fig.~\ref{Exp_data}(a). Here the deviation between
the theoretical curve and the experimental data is never larger than
$2-3$ k$\Omega/\mu$m. Treating the electrons as non-interacting, on
the other hand, gives a deviation which increases with temperature and
is of the order of $7$ k$\Omega/\mu$m at $T=300$ K. Qualitatively
however, $\rho(T)\sim T^{(1+K_{c+})/2}$, in the approximation used and
with $K_{c+}<1$, gives a concave function while the experimental curve
is convex. In the zero temperature limit, the theoretically calculated
$\rho(T)$ then has, in particular, infinite slope contrary to what is
observed experimentally. Here it is, however, important to note that
our approximation, using an infinite wire to calculate $\gamma$,
breaks down for $\xi_e\gtrsim L$, i.e., $T<T_c\approx 30$ K. In the
limit $T\to 0$ the properties will be instead dominated by the leads
and we have to essentially use the free electron Green's function in
(\ref{Self-E}) so that the theoretical curve has to smoothly connect
to the non-interacting result in this limit.

The relevant microscopic parameters such as the electron-phonon
coupling $g_2$ or the Luttinger parameter $K_{c+}$ are only
approximately known while the chiral angle $\eta$ is even completely
undetermined. We can thus try to improve the agreement with experiment
by varying these parameters within a reasonable range. The
measurements of the tunnel conductance\cite{BockrathCobden,YaoPostma}
seem to be convincing evidence that single wall carbon nanotubes are
Luttinger liquids with $K_{c+}\approx 0.2-0.4$. In the experiment
considered here this nanotube sits on an insulating substrate so there
is no reason to expect that $K_{c+}$ is significantly changed by
screening effects. For the coupling constant $g_2$, on the other hand,
estimates obtained from experimental data and theoretical calculations
vary between $g_2\sim 1-3$
eV.\cite{Hertel2000,Pietronero1980,MartinoEgger} Since $g_2$ enters
quadratically into the resistivity formulas this means that we can
vary $\rho(T)$ by almost an order of magnitude. A variation of the
chiral angle $\eta$ for a tube with fixed diameter has a substantial
effect on the resistivity as well as we have already demonstrated in
Fig.~\ref{Exp_data}(a,b). However, as long as we keep $K_{c+}<1$ the
temperature dependence will never fully agree with the experimentally
observed one.

Finally, we might try to fit the experimental data by assuming that
the acoustic modes are quenched and the resistivity is caused purely
by a coupling to the breathing mode. 
\begin{figure}
\includegraphics*[width=1.0\columnwidth]{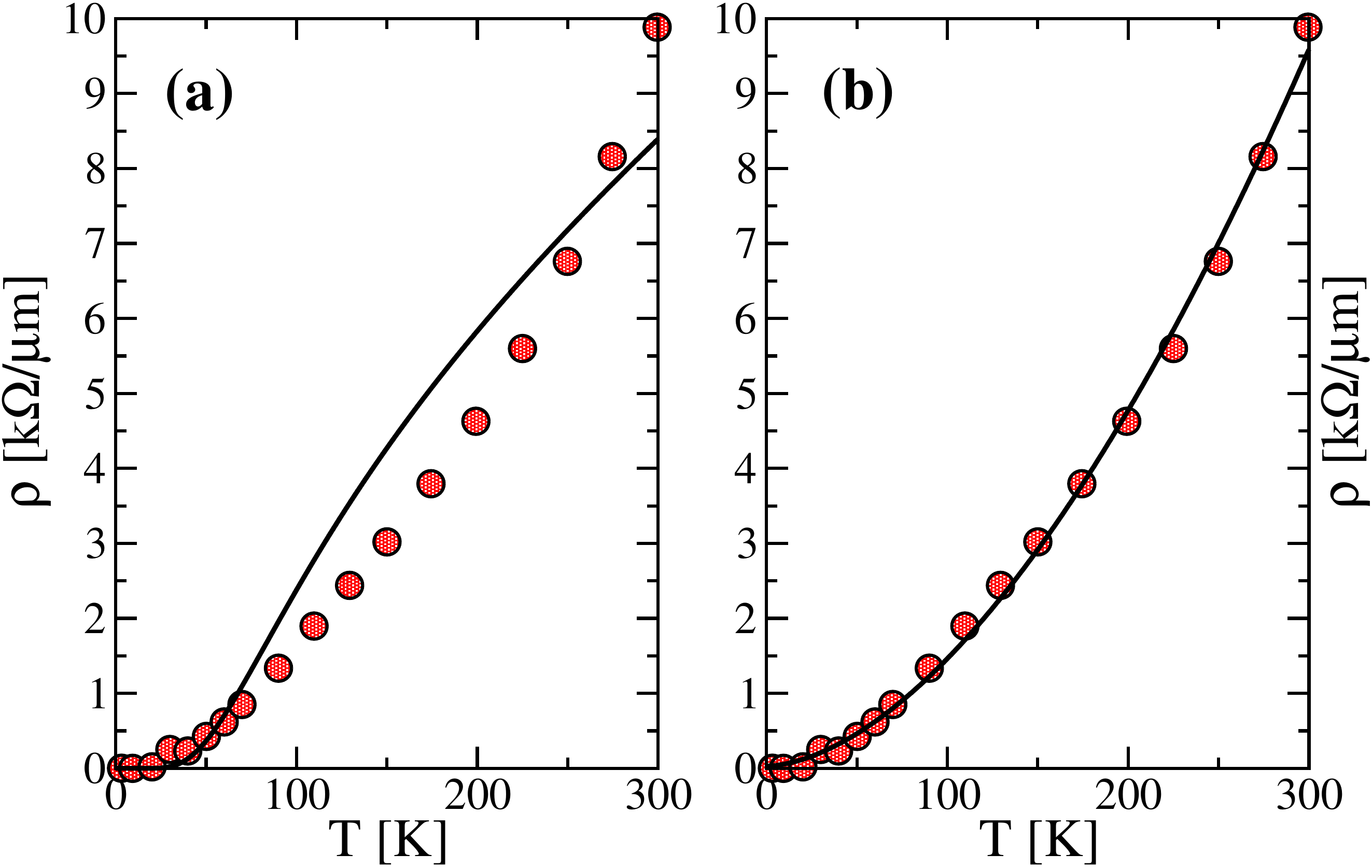}
\caption{(Color online) Experimental data taken from
  Ref.~\onlinecite{PurewalHong} (dots) compared to: (a) the
  theoretical results for the breathing mode of a $(27,0)$ zigzag tube
  assuming $K_{c+}=0.4$, $g_2=4.5$ eV, and $\hbar\omega_B/k_B=250$ K.
  (b) a fit to a power law yielding
  $\rho(T)=2770\frac{\text{k}\Omega}{\mu\text{m}}(\pi ak_B
  T/v_F\hbar)^{1.73}$.}
\label{Exp_data2}
\end{figure}
The result of such a fit is shown in Fig.~\ref{Exp_data2}(a) but the
electron-phonon coupling $g_2=4.5$ eV needed is substantially larger
than the largest estimates. This scenario thus seems unrealistic and
the fit is not too convincing either. On the other hand, we can obtain
an excellent fit of the data over the whole temperature range with a
single power law as shown in Fig.~\ref{Exp_data2}(b). The exponent
obtained in this fit is larger by about $1$ compared to the exponent
obtained from the theoretical calculation at $T\gg T_c$ for the
twiston and stretching contributions. Interestingly, we can obtain
such a power law where even the prefactor is of the right magnitude by
multiplying our result (\ref{twiston_relaxation}) for the resistivity
due to the twiston mode with the characteristic phonon scale,
$\rho(T)\to \rho(T)\, k_B T a/(v_t\hbar)$. While this could be
accidental, the fact that a single power law yields an excellent fit
over a wide temperature range and that the damping due to the phononic
degrees of the tube gives a resistivity of the right magnitude is a
strong indication that this is indeed the dominant mechanism although
the theory so far cannot fully explain the temperature scaling. We
will speculate about possible modifications of this theory and discuss
alternative explanations which can be found in the literature in the
conclusions.

\section{Conclusions}
\label{Sec:Concl}
The purpose of this paper has been twofold: On the one hand, we wanted
to derive a general formula for the conductance of an interacting
quantum wire with good contacts and current relaxing processes in the
wire. On the other hand, we wanted to qualitatively understand the
temperature and length-dependent scaling of the conductance measured
in a recent experiment\cite{PurewalHong} on single wall carbon
nanotubes.

Concerning the first part, we have shown that the approach by Maslov
and Stone\cite{MaslovStone} for an interacting ballistic wire
contacted to leads can be generalized to an interacting wire with
damping. Under the assumptions that a Luttinger liquid description is
valid and that the current relaxing processes are predominantly taking
part in the bulk of the wire, we were able to give the result for the
electronic Green's function for such a setup in closed form. This
allowed us, in particular, to calculate the dc resistance which turned
out to consist of two terms. The first is the quantum resistance
multiplied by the Luttinger parameter $K_\ell$ of the leads. As in the
case without damping, there is thus no renormalization if we assume
Fermi liquid leads, $K_\ell=1$. The second term is proportional to the
length of the wire and the damping rate $\gamma$. While this general
structure seems obvious, the important point is that $\gamma$, in this
approximation, depends only on the Luttinger parameter in the wire
$K_{\mathrm{w}}$. We thus have a separation into a constant term, the
quantum resistance, which only depends on the properties of the leads,
and a term proportional to the length which only depends on the
properties of the wire.

As a first application, we have calculated the resistance of an
interacting quantum wire which has coexisting ballistic and diffusive
channels. Such a coexistence is expected for integrable models where
part of the current is protected by a local or quasi-local
conservation law.\cite{SirkerPereira2,Prosen} We find that in such a
case the ballistic channel, however small, completely dominates the
transport so that the system still shows ideal quantum conductance.
However, it is important to mention that two assumptions have been
made which cannot be fulfilled in practice: (1) Attaching an
integrable wire to contacts will, in general, lead to non-integrable
boundary conditions. (2) There will also be relevant backscattering at
the contacts which we have neglected. Nevertheless, what this
calculation does show is that a quantum wire which has an almost
ballistic transport channel can display a conductance which stays
close to the ideal quantum conductance over a wide temperature range.

In the second part of the paper we have calculated the resistance of
single-wall carbon nanotubes caused by a coupling to the phononic
degrees of freedom of the tube. It is well known that there are three
modes which have to be taken into
account.\cite{SuzuuraAndo,MartinoEgger} These are an acoustic twist-
and stretching mode as well as an optical breathing mode. Previously,
only the contribution of the twiston mode, which is the only one of
the three modes active for an armchair tube, had been considered. If
one assumes that the electrons are non-interacting then the coupling
of the electrons to the twist distortions of the tube leads to a
resistivity which increases linearly with
temperature.\cite{KaneMeleLee} Interactions, however, are expected to
change this temperature dependence.\cite{KomnikEgger} In our paper we
have calculated the contribution of all three modes by using a
Luttinger liquid theory for the electrons and including the
electron-phonon coupling in terms of a self-energy---obtained in
second order perturbation theory---for the current-current correlation
function. The result depends only on microscopic parameters of the CNT
which mostly are relatively well known. A comparison with experiment
shows that our results, which do take the electron-electron
interactions into account, agree quantitatively better with experiment
than the formula for non-interacting electrons. However, qualitatively
the observed temperature dependence of the resistivity is different
from the calculated one. For temperatures $T$ such that the coherence
length $\xi_e\ll L$ we find $\rho\sim T^\alpha$ with an exponent
$\alpha=(1+K_{c+})/2$ determined by the Luttinger parameter of the
total charge mode which theoretically and experimentally has been
estimated to be around
$K_{c+}=0.2-0.4$\cite{EggerGogolin,BockrathCobden,YaoPostma} leading
to $\alpha=0.6-0.7$. The experimental data, on the other hand, are
described extremely well by a power law with exponent $\alpha\approx
1.7$. 

Let us discuss possible reasons for these deviations and alternative
explanations in the following. First, there is no reason to expect
that $K_{c+}$ substantially deviates from the used value. In the
experiment the tube is placed on an insulating substrate so that
screening should not take place. Interactions with phonons can lead to
a renormalization of $K_{c+}$ but are only effective if the Coulomb
interaction is already screened.\cite{MartinoEgger} A renormalization
to values leading to the observed exponent $\alpha\approx 1.7$ is out
of the question in any case. Since Luttinger liquid properties have
been theoretically predicted\cite{EggerGogolin} and experimentally
observed\cite{BockrathCobden,YaoPostma} there seems to be no good
reason to doubt the electronic part of our theory. This leaves the
phononic part. There is a damping of the phonons due to phonon-phonon
interactions which will modify the phonon
propagator.\cite{MartinoEggerGogolin} However, this does not have any
effect on our calculations since we are always effectively in a high
temperature regime for the acoustic phonons because the sound
velocities of the two acoustic modes are much smaller than the Fermi
velocity. The phonon modes are therefore always heavily populated and
the phonon propagator simply becomes $D(q,\tau)\propto T$ to leading
order even if damping is included.

In several papers the possibility of a Peierls distortion of the CNT
has been discussed.\cite{SedekiCaron,MintmireDunlap,FiggeMostovoy} If
we have a distortion which leads to a finite expectation value of one
of the acoustic modes, $\langle\partial_x\Phi\rangle\neq 0$ in
Eq.~(\ref{twist-coup}), then we can easily obtain the temperature
dependence of the resistivity in the perturbative regime by scaling
arguments. To obtain the self-energy in this case we now have to
integrate over time {\it and} space and there is a factor of $T$
missing which before came from the phonon propagator. The scaling is
therefore given by $\rho\sim T^{(K_{c+}-3)/2}$ in the perturbative
regime while $\rho(T)$ will show thermally activated behavior at very
low temperatures in the presence of a Peierls distortion. None of this
is seen in the experimental data.

If the tube is exactly at half-filling then purely electronic Umklapp
scattering is also possible.\cite{BalentsFisher} Note that the same
term is called a forward scattering term in
Ref.~\onlinecite{EggerGogolin}. The three Umklapp operators in
bosonized form are given by $\hat O_U\sim
\cos(4q_Fx+2\sqrt{\pi}\phi_{c+})\cos(2\sqrt{\pi}\phi_{j\delta})$ with
$j\delta\neq c+$. Calculating again the self-energy for such a process
in the perturbative regime and using scaling arguments we find that
the resistivity due to Umklapp scattering scales as $\rho\sim
T^{2K_{c+}-1}$. In particular, the temperature dependence is linear in
the non-interacting case, $K_{c+}=1$, as already stated in
Ref.~\onlinecite{BalentsFisher}. At very low temperatures Umklapp
scattering at half-filling will lead to gaps both in the charge and in
the spin sector and thus to thermally activated behavior for
$\rho(T)$. In a device configuration the filling in the tube is,
however, usually tuned away from half-filling so that the Umklapp term
oscillates with $4q_F$ and can thus be neglected at temperatures
$k_BT\ll\hbar v_F q_F$. Even if Umklapp scattering does contribute at
higher temperatures its contribution will be smaller by a factor
$R_a^{-1}$ compared to the electron-phonon contribution so that for
the tubes considered here the latter will always be dominant.

There is an alternative explanation for the resistivity of CNTs on
surfaces which can be found in the
literature.\cite{ChandraPerebeinos,PerebeinosRotkin} According to
these papers the main contribution to the resistivity at room
temperature stems from a coupling of the electrons to surface modes of
the substrate. The experimental data are explained by combining the
resistivity due to the coupling to the acoustic modes of the tube,
which dominates at low temperatures, with the resistivity stemming
from a coupling to the optical surface modes which yield the main
contribution at higher temperatures. In the calculation of both
contributions the electrons are treated as non-interacting. This
assumption seems questionable. Our calculation shows that once
electron-electron interactions are included the interactions with the
phonon modes of the tube alone give a resistivity of the right
magnitude even at room temperature if standard parameters for the tube
are used. If surface modes do indeed give the dominant contribution at
room temperature then this would therefore require that the
contribution due to the phonon modes of the tube is much smaller than
expected. In particular, such a scenario requires $g_2\lesssim 0.5$
eV.

While our results provide a much better description of the
experimental data than a non-interacting theory, the observed
deviations have to remain as an interesting unsolved puzzle. A first
step to resolve it would be experiments on free-standing tubes to see
if the resistance substantially changes compared to the ones with the
tube on a substrate which we considered here.

\begin{acknowledgments}
  The authors thank P.~Kim for sending us the experimental data of
  Ref.~\onlinecite{PurewalHong} and I.~Affleck and R.~Egger for
  valuable discussions. J.S.~thanks the Galileo Galilei Institute for
  Theoretical Physics for their hospitality and the INFN for partial
  support. We also acknowledge support by the DFG via the SFB/TR 49
  and by the graduate school of excellence MAINZ.
\end{acknowledgments}

\appendix 
\section{Green's function for a quantum wire with contacts and damping}
\label{App_GF}
To calculate the Green's function for the action (\ref{Action}) the
equation (\ref{GF_Eq1}) together with the boundary conditions have to be solved.
There is a discontinuity in the derivative of $g_m$ at $x=x'$:
\begin{eqnarray}
\label{GF_Eq2}
\frac{v_x}{K_x}\partial_xg_m(x,x')|_{x=x'-0}^{x=x'+0}=1.
\end{eqnarray}
For the boundaries at $y=0,L$ we have furthermore
\begin{eqnarray}
\label{GF_Eq3}
\frac{v_x}{K_x}\partial_xg_m(x,x')|_{x=y-0}^{x=y+0}=0.
\end{eqnarray}
where $x\neq x'$. $g_m(x,x')$ itself must be continuous everywhere.
We have also fixed $\lim_{x\to\pm\infty}g_m(x,x')\to0$. Note that
$g_m(x,x')$ must be symmetric under swapping $x$ and $x'$.  For $x$ in
the three different regions and focusing on $0<x'<L$ we make the
ansatz
\begin{eqnarray}
\label{GF_ansatz}
g_m(x,x')=\left\{
\begin{array}{cc}
A\e^{\,\frac{|\omega_{m}|x}{v_\ell}} &x<0\\B\e^{\frac{\omega_{m\gamma}x}{v_{\mathrm{w}}}}+C\e^{-\frac{\omega_{m\gamma}x}{v_{\mathrm{w}}}}&0<x<x'\\D\e^{\frac{\omega_{m\gamma}x}{v_{\mathrm{w}}}}+E\e^{-\frac{\omega_{m\gamma}x}{v_{\mathrm{w}}}}&x'<x<L\\F\e^{-\frac{|\omega_m|x}{v_\ell}}&x>L
\end{array}
\right.
\end{eqnarray}
where $\omega_{m\gamma}=\sqrt{\omega^2_m+2\gamma K_{\mathrm{w}}|\omega_m|v_{\mathrm{w}}}$ for the damped case and $\omega_{m\gamma}=\sqrt{(1+K_{\mathrm{w}}v_{\mathrm{w}}y^{-1})\omega_m^2}$ for the the protected current case.

The boundary conditions then lead to the set of equations
\begin{subequations}
\begin{eqnarray}
A&=&B+C\\
B-D&=&e^{-2\frac{x'\omega_{m\gamma}}{v_{\mathrm{w}}}}(E-C)\\
Fe^{-\frac{|\omega_m|L}{v_\ell}}&=&De^{\frac{L\omega_{m\gamma}}{v_{\mathrm{w}}}}+Ee^{-\frac{L\omega_{m\gamma}}{v_{\mathrm{w}}}}\\
B-C&=&A\frac{K_{\mathrm{w}}}{K_\ell}\frac{|\omega_m|}{\omega_{m\gamma}}\\
Fe^{-\frac{|\omega_m|L}{v_\ell}}&=&\frac{K_\ell}{K_{\mathrm{w}}}\frac{\omega_{m\gamma}}{|\omega_m|}
\left[Ee^{-\frac{L\omega_{m\gamma}}{v_{\mathrm{w}}}}-De^{\frac{L\omega_{m\gamma}}{v_{\mathrm{w}}}}\right]\\
B-D&=&-\frac{K_{\mathrm{w}}e^{-\frac{x'\omega_{m\gamma}}{v_{\mathrm{w}}}}}{2\omega_{m\gamma}}
\end{eqnarray}
\end{subequations}
For convenience we will define
\begin{eqnarray}
\label{barK}
\bar{K}_\omega=\frac{K_\ell\omega_{m\gamma}+K_{\mathrm{w}}|\omega_m|}{K_\ell\omega_{m\gamma}-K_{\mathrm{w}}|\omega_m|}.
\end{eqnarray}

This set of equations is readily solved and one finds
\begin{subequations}
\begin{eqnarray}
B&=&\frac{K_{\mathrm{w}} \bar{K}_{\omega} }{2\omega_{m\gamma}}\frac{e^{\frac{x' \omega_{m\gamma}}{v_{\mathrm{w}}}}+e^{\frac{(2 L-x') \omega_{m\gamma}}{v_{\mathrm{w}}}} \bar{K}_{\omega}}{1-e^{\frac{2 L \omega_{m\gamma}}{v_{\mathrm{w}}}}\bar{K}_{\omega}^2}\\
C&=&\frac{K_{\mathrm{w}} }{2\omega_{m\gamma}}\frac{e^{\frac{x' \omega_{m\gamma}}{v_{\mathrm{w}}}}+e^{\frac{(2 L-x') \omega_{m\gamma}}{v_{\mathrm{w}}}} \bar{K}_{\omega}}{1-e^{\frac{2 L \omega_{m\gamma}}{v_{\mathrm{w}}}}\bar{K}_{\omega}^2}\\
D&=&\frac{K_{\mathrm{w}} }{2\omega_{m\gamma}}\frac{e^{-\frac{x' \omega_{m\gamma}}{v_{\mathrm{w}}}}+e^{\frac{x' \omega_{m\gamma}}{v_{\mathrm{w}}}} \bar{K}_{\omega}}{1-e^{\frac{2 L \omega_{m\gamma}}{v_{\mathrm{w}}}}\bar{K}_{\omega}^2}\\
E&=&\frac{K_{\mathrm{w}} \bar{K}_{\omega} }{2\omega_{m\gamma}}\frac{e^{-\frac{x' \omega_{m\gamma}}{v_{\mathrm{w}}}}+e^{\frac{x' \omega_{m\gamma}}{v_{\mathrm{w}}}} \bar{K}_{\omega}}{e^{-\frac{2 L \omega_{m\gamma}}{v_{\mathrm{w}}}}-\bar{K}_{\omega}^2}.
\end{eqnarray}
\end{subequations}
This completes the determination of the parameters in
(\ref{GF_ansatz}) which is thus the full bosonic Green's function for
a quantum wire with contacts and damping in the bulk.

Analytic continuation gives us $\omega_m\to-i\omega+\delta$, and defining $\omega_{\gamma}=i\omega_{m\gamma}|_{\omega_m=-i\omega}$ and $\bar{K}_{\omega}=\bar{K}_{\omega_m=-i\omega}$ we have
\begin{eqnarray}
\omega_{\gamma}&=&\left\{
\begin{array}{cc}
\sqrt{\omega^2+2i\gamma K_{\mathrm{w}}\omega v_{\mathrm{w}}} & \textrm{ or}\\*[0.1cm]
\sqrt{(1+K_{\mathrm{w}}v_{\mathrm{w}}y^{-1})\omega^2} &
\end{array}\right.
\end{eqnarray}
where the first line applies to the damped case with self energy
$\Sigma\approx -2\im\gamma\omega$ and the second line to the case
where part of the current is conserved with a self energy
$\Sigma\approx -y^{-1}\omega^2$. In addition, we now have
\begin{eqnarray}
\bar{K}_{\omega}&=&\frac{K_\ell\omega_{\gamma}+K_{\mathrm{w}}\omega}{K_\ell\omega_{\gamma}-K_{\mathrm{w}}\omega}.
\end{eqnarray}

In the low-frequency limit we note that the Green's function becomes
position independent:
\begin{eqnarray}
\label{Green_app}
g(x,x';\omega)\approx\frac{iK_{\mathrm{w}}\left(1+\bar{K}_{\omega}\right)}{2\omega_\gamma}\frac{1+e^{-\im\frac{ 2 L \omega_{\gamma}}{v_{\mathrm{w}}}}\bar{K}_{\omega}}{1-e^{-\im\frac{ 2 L \omega_{\gamma}}{v_{\mathrm{w}}}}\bar{K}_{\omega}^2}.
\end{eqnarray}
Now, for the damped case as $\omega\ll\gamma K_{\mathrm{w}}v_{\mathrm{w}}$ we have
$\omega_{\gamma}\sim e^{-i\pi/4}\sqrt{2\gamma K_{\mathrm{w}}\omega v_{\mathrm{w}}}$.
In the protected current scenario we find in the low-frequency limit
\begin{equation}
g(x,x';\omega)\approx\frac{iK_{\mathrm{w}}}{2\omega\sqrt{(1+K_{\mathrm{w}}v_{\mathrm{w}}y^{-1})}}\frac{1+\bar{K}_{\omega}}{1-\bar{K}_{\omega}}\, . 
\end{equation}

\section{Parameters for carbon nanotubes}
\label{App}
In Table \ref{CNT_parameters}, we list all the relevant microscopic
parameters to obtain the resistance of single-wall carbon nanotubes
caused by electron-phonon coupling. For some parameters significantly
varying estimates have been given in the literature in which cases we
give the range of these estimates.

\begin{table*}
\caption{\label{CNT_parameters} Relevant parameters for the resistance of carbon nanotubes}
\begin{ruledtabular}
\begin{tabular}{lll}
Quantity & Symbol & Value\\
\hline\\
Lattice constant & $a$ & $2.46\cdot 10^{-10}$ m\\
Carbon mass per unit area & $M$ & $3.8\cdot 10^{-7}$ kg/$\text{m}^2$\\
Fermi velocity & $v_F$ & $8\cdot 10^5$ m/s\\
Radius of a $(n,m)$ tube & $R_a$ & $\frac{a}{2\pi}\sqrt{n^2 + nm + m^2}$\\
Bulk modulus\cite{Dolling1962} & $B$ & $110.2$ kg/$\text{s}^2$\\
Shear modulus\cite{Dolling1962} & $\mu$ & $57.38$ kg/$\text{s}^2$\\
Twiston velocity\cite{SuzuuraAndo} & $v_t$ & $1.23\cdot 10^4$ m/s \\
Twist modulus for $(n,n)$ tube\cite{KaneMeleLee} & $C_t$ & $n^3 18\cdot 10^{-10}$ eV m\\
Twist moment of inertia per unit length & $M_t$ & $C_t/v_t^2$  [kg m]\\
Stretching mode velocity\cite{SuzuuraAndo} & $v_s$ & $1.99\cdot 10^4$ m/s\\
Breathing mode frequency\cite{SuzuuraAndo} & $w_B$ & $ \sqrt{\frac{B+\mu}{M}}\frac{1}{R_a}$ 1/s\\
Factor between twiston and stretching propagator & $\mathcal{A}$ & $\frac{4B-Mv_s^2}{B +\mu}\l(\frac{v_t}{v_s}\r)^2\approx 0.66$\\
Quantum resistance & $R_Q=\frac{h}{4e^2}$ & $6.45$ k$\Omega$\\
Transfer integral\cite{SuzuuraAndo} & $\gamma_0$ & $3.0$ eV\\
Transfer integral & $\beta$ & $1.1$ eV\cite{Hertel2000}, $3.6$ eV\cite{Pietronero1980}\\
Electron-phonon coupling constant\cite{SuzuuraAndo} & $g_2=\frac{3\mu\beta\gamma_0}{4\sqrt{2}B}$ & $0.91 - 2.98$ eV; $1.5$ eV\cite{MartinoEgger}\\
\end{tabular}
\end{ruledtabular}
\end{table*}


\end{document}